\documentclass{osa-article}

\journal{osajournal}

\articletype{Research Article}

\begin{document}

\title{Prescription AR: A Fully-Customized Prescription-Embedded Augmented Reality Display}


\author{Jui-Yi Wu and Jonghyun Kim\authormark{*}}
\address{NVIDIA}

\email{\authormark{*}jonghyunk@nvidia.com} 


\begin{abstract}
In this paper, we present a fully-customized AR display design that considers the user's prescription, interpupillary distance, and taste of fashion. A free-form image combiner embedded inside the prescription lens provides augmented images onto the vision-corrected real world. We establish a prescription-embedded AR display optical design method as well as the customization method for individual users. Our design can cover myopia, hyperopia, astigmatism, and presbyopia, and allows the eye-contact interaction with privacy protection. A 169$g$ dynamic prototype showed a 40$^\circ$ $\times$ 20 $^\circ$ virtual image with a 23 cpd resolution at center field and 6 mm $\times$ 4 mm eye box, with the vision-correction and varifocal (0.5-3$m$) capability.
\end{abstract}

\newcommand{\juiyi}[1]{{\color{blue}{JUIYI: #1}}}
\newcommand{\jonghyun}[1]{{\color{red}{JONGHYUN: #1}}}
\newcommand{\Pete}[1]{{\color{red}{PETE: #1}}}
\newcommand{\Kaan}[1]{{\color{red}{KAAN: #1}}}
\newcommand{\Dave}[1]{{\color{red}{DAVE: #1}}}
\newcommand{\Morgan}[1]{{\color{red}{MORGAN: #1}}}
\newcommand{\Josef}[1]{{\color{red}{JOSEF: #1}}}
\newcommand{\Ben}[1]{{\color{red}{BEN: #1}}}
\newcommand{\Joohwan}[1]{{\color{red}{JOOHWAN: #1}}}
\newcommand{\Rachel}[1]{{\color{red}{RACHEL: #1}}}
\newcommand{\Mark}[1]{{\color{red}{Mark: #1}}}
\newcommand{\Zander}[1]{{\color{red}{Zander: #1}}}
\newcommand{\Ward}[1]{{\color{red}{Ward: #1}}}
\newcommand{\Michael}[1]{{\color{red}{Michael: #1}}}

\section{Introduction}
\label{sec:Introduction}
Augmented Reality (AR) displays present virtual images at real-world scenes while preserving the viewer's natural vision. Numerous AR head mounted displays (HMDs) have been introduced in both commercial prototypes\cite{googleglasses,Hololens1,Hololens2,DIGILENS,Lumus} and research literature\cite{akcsit2017near,maimone2017holographic,maimone2014pinlight,lee2017foveated, kim2019foveated} since Google glass, but no device has achieved wide adoption by consumers. In contrast to market expectations on AR displays represented by "holographic" images, there is no viable optical structure meeting both of the slim form factor and high visual standards including high-resolution, large field-of-view (FOV), large eye-box, and variable focus. 

The diversity of human head shape and eye structure aggravates this challenge further. Every user has different interpupillary distance (IPD, 50 - 75 mm)\cite{dodgson2004variation} and nose shape, which raises the bar on eye box and eye relief coverage beyond the requirement for a single user. In addition, since more than 40\% of the population uses special aids for vision correction caused by myopia, hyperopia, astigmatism, and presbyopia, AR display should consider the viewer's prescription\cite{holden2016global,pan2015age,hashemi2018global}. AR display manufacturers produced either an eye-glasses compatible design or an additional prescription-lens option, but both methods increase the weight and the size of the system significantly.

Several approaches have been introduced to include vision correction in an AR display structure\cite{maimone2017holographic,NorthAR,wang2017augmented,zhou2017see}. The North AR demonstrated the most ergonomic eye-glasses like AR displays by measuring customer's prescription and IPD~\cite{NorthAR}. However, previous methods were mostly based on the retinal projection display using holographic optical elements (HOEs) attached to the rear surface of the prescription lens, which inherently causes limited FOV at given eye relief and narrow eye box\cite{NorthK,zhou2017see}. Considerable dynamic eye box expansion methods have been proposed\cite{jang2018holographic,jang2017retinal}, but they required additional a spatial light modulator or a linear actuator and are not suitable for an eyeglasses form factor. Above all, we are not aware of any previous customized AR display design which considers the observer's prescription, IPD, and facial appearance.

In this paper, we propose a fully-customized, prescription-embedded AR display. We propose an optical design that utilizes the prescription lens as a wave-guide for the AR display. A free-form image combiner is inserted into the prescription lens, so one lens piece can deliver virtual scenes and correct the vision of the real scene simultaneously. Based on a modified myopia eye model, we first design the shape of the prescription lens. Then the free-form image combiner, in-coupling prism, and beam shaping lens are optimized for the individual prescription lens. In addition, a customized ergonomic eye-glasses design is achieved by using a 3D facial scan. A Prescription AR prototype with a 5-mm thick lens provides 1 diopter (1D) vision correction, 23 cycles per degree (CPD) angular resolution at center, 6 mm eye box, and varifocal (0.33D - 2D) capability. The prototype is lightweight (169$g$ for dynamic and 79$g$ for static prototype), has 70\% transparency, protects user's privacy, and enables eye-contact interaction with surroundings.

\section{Design}
\label{sec:Design}
\subsection{System overview}
\label{sec:system_overview}
\begin{figure}[hhhh]
\centering
\includegraphics[width=5in]{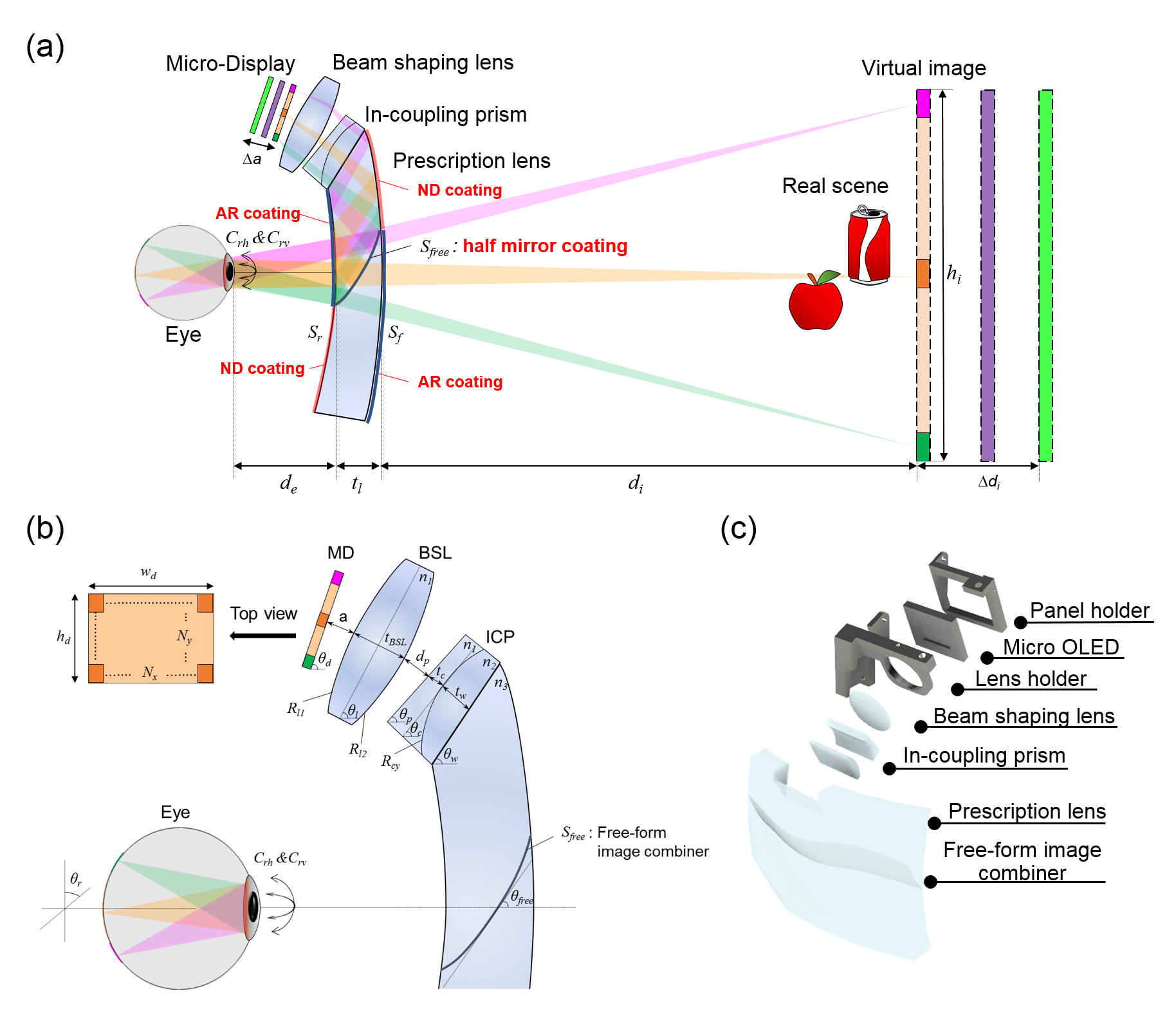}
\caption{\textmd{Schematic diagram of Prescription embedded AR display: (a) The side view and the beam path of the AR image of the proposed system. The prescription lens works both for vision-correction and for wave-guide of the AR image. Light rays from a micro display (MD) refracted by a beam shaping lens (BSL) enter to the prescription lens (PL) through an in-coupling prism (ICP) and create a magnified virtual image located a distance $d_i$ from the eye. The focal image depth can be dynamically changed from 0D to 2D by moving MOLED axially,$\Delta a$, in experimental results. (b) The Detailed diagram for geometric parameters in the prescription embedded AR display. (c) The 3D diagram of optical components.}}
\label{fig:schematic diagram of prescription AR}
\end{figure}

As its core, our prescription AR display optically corrects the user's vision with a prescription lens and utilizes that lens as a wave-guide in an AR display system. As shown in Fig. \ref{fig:schematic diagram of prescription AR}(a), the top surface of the prescription lens of thickness $t_l$ is used as an entrance of the wave-guide. The light rays from a micro display of size $w_d$$\times$$h_d$ and resolution $N_{x}$$\times$$N_{y}$ located in front of user's forehead with an angle $\theta_d$ are refracted by a bi-convex ($R_{l1}$, $R_{l2}$) beam shaping lens (refractive index: $n_1$, thickness: $t_{BSL}$) located at $a$ from the micro display with the tilted angle $\theta_l$ and entered to the wave-guide through an in-coupling prism (refractive index: $n_1$) located at $d_p$ from the beam shaping lens with the tilted angle $\theta_p$. The in-coupling prism is composed by a set of a plano-concave and a convex-plano cylindrical lens. Then, the rays are refracted by a cylindrical lens ($R_{cy}$, refractive index: $n_2$) located at $t_c$ from the prism surface with the tilted angle $\theta_c$ and travel in the wave-guide (refractive index: $n_3$, tilted angle: $\theta_w$) located at $t_w$ from the cylindrical surface as shown in Fig. \ref{fig:schematic diagram of prescription AR}(c). The light rays are total internal reflected (TIR) twice by the frontal surface ($S_f$) and the rear surface ($S_r$) of the prescription lens, reflected by a free-form half-mirror coated surface ($S_{free}$, tilted angle: $\theta_f$), and arrived at the pupil of the eye. Note that the in-coupling prism, cylindrical lens, upper part of the prescription lens, and lower part of the prescription lens are bonded by an optical adhesive, so the prototype consists of only two lens pieces: the main lens and the beam shaping lens. The enlarged virtual image of size $w_i$$\times$$h_i$ is located at distance $d_i$ from the eye in the vision-corrected real scene. The virtual image depth can be dynamically adjusted ($\Delta d_i$) by moving the micro display back and forth ($\Delta a$).

\begin{figure}[hhhh]
\centering
\includegraphics[width=5in]{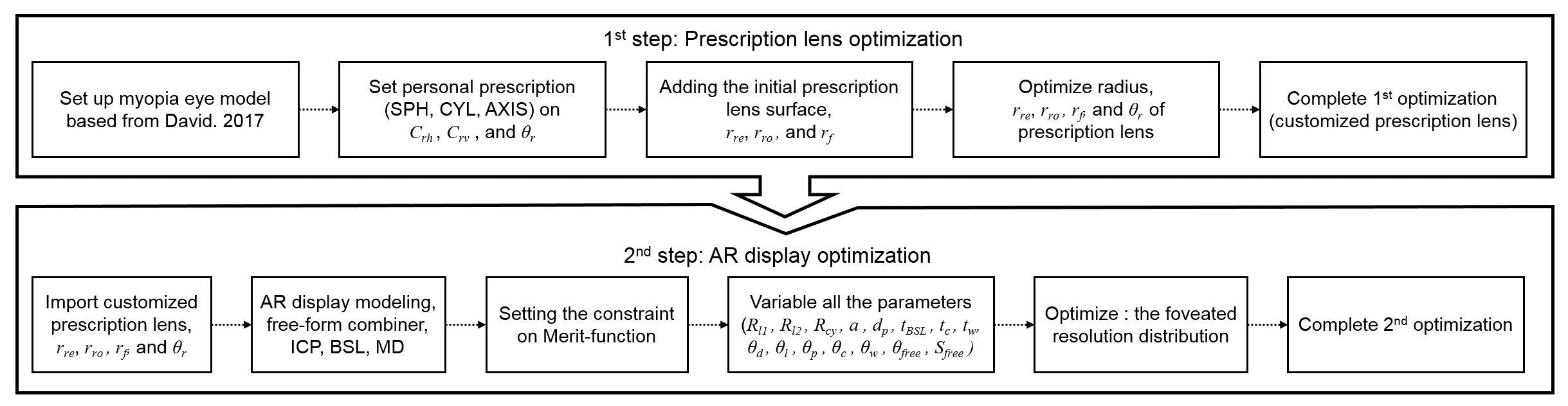}
\caption{\textmd{Flow chart of the two-step optimization for Prescription AR: The frontal and rear surfaces ($S_f$, $S_r$) are optimized at given lens thickness $t_l$ in the first step based on user's prescription. Other geometric parameters ($R_{l1}$, $R_{l2}$, $R_{cy}$, $a$, $d_p$, $t_{BSL}$, $t_c$, $t_w$, $\theta_d$, $\theta_l$, $\theta_p$, $\theta_c$, $\theta_w$, and $\theta_{free}$) are optimized in the second step based on target foveated resolution and eye relief range,$d_e$.}}
\label{fig:Flow chart of prescription embedded AR display}
\end{figure}

The optical design process is divided into two steps: the prescription lens design ($S_f$, $S_r$) and the AR display path design (rest of all) using a commercial optics simulation tool, Zemax OpticStudio. The overall optical path is infeasible to be investigated by an analytic form because of the free-form surface and the multiple off-axis components. Nevertheless, we propose a universal designing and optimization method which is valid for any kind of prescription including myopia, astigmatism, hyperopia, and presbyopia. Figure \ref{fig:Flow chart of prescription embedded AR display} shows a flow chart of the universal 2-step optimization process in prescription AR, which is started from the user's eyeglasses prescription including spherical correction(SPH), cylinder correction(CYL), axis of astigmatism(AXIS), and add power (ADD).

\subsection{Prescription lens design from modified myopia eye model}
\label{sec:prescription_lens_design}
\begin{figure}[htb]
\centering
\includegraphics[width=4.5in]{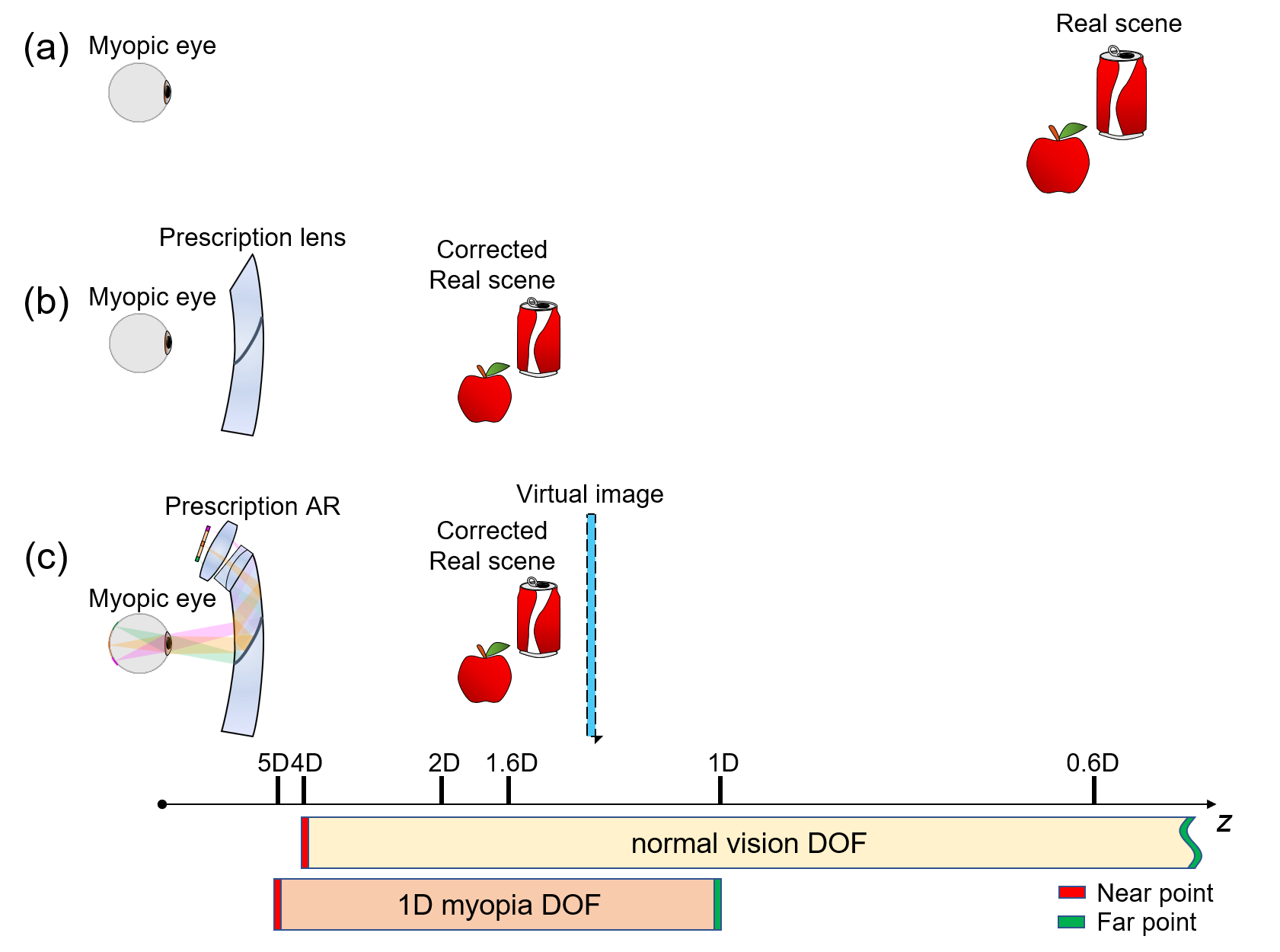}
\caption{\textmd{Principle of the Prescription AR: (a) A 1D myopic eye and its accommodation range of 1D to 5D. The 1D myopic eye cannot resolve the real objects located at 0.6D. (b)The prescription lens design for 1D myopic eye. The lens effectively shifts the object from infinity (0D) to the myopia's far point (1D). The object located at 0.6D is imaged at 1.6D plane with the correct prescription lens. (c) The prescription AR design. The virtual image plane should be located inside the accommodation range of the myopic eye. Note that the depth of field (DOF, accommodation range) of normal vision is from 0D to 4D and the DOF of $X$D myopic eye is from $X$D to ($X+4$)D.}}
\label{fig:Prescription_AR_design}
\end{figure}

The first step is the optimization of the prescription lens, especially the frontal ($S_f$) and rear ($S_r$) surface profile of the prescription lens\cite{sun2000ophthalmic,Ophthalmiclens}. Figure \ref{fig:Prescription_AR_design} shows how to design the prescription lens for myopic eye. Figures \ref{fig:Prescription_AR_design}(a) and \ref{fig:Prescription_AR_design}(b) show the principle of vision correction for myopic eye. The normal vision whose amplitude of accommodation is 4D has a far point at 0D and near point at 4D. The 1D myopic eye with the same amplitude of accommodation has a far point at 1D and near point at 5D. The observer cannot perceive full-resolution image of the objects located at 0.6D because the object is out of accommodation range, as shown in Fig. \ref{fig:Prescription_AR_design}(a). The prescription lens shifts the object at infinity to the myopic eye's far point, 1D location, so the objects are imaged at 1.6D plane, inside the accommodation range of 1D myopic eye. Similarly, the prescription AR display offer the clear enlarged virtual image inside the accommodation range of myopic eye. In additionally, the prescription lens compensates the astigmatism, by adding inverse cylinder power to the given axis.

Instead of using the direct calculation of surface profiles from the SPH, CYL, AXIS, and ADD values, we optimized both surfaces using a human eye model. This optimization method can minimize the aberration at the given thickness $t_l$, refractive index $n_3$, and given eye relief $d_e$. Previously, Atchison built human myopic eye model based on the measured data from 121 subjects \cite{atchison2006optical}. And it is known that the total astigmatism is the sum of the corneal and internal astigmatism\cite{hoffmann2010analysis}. However, there isn't a general human eye model covering both myopia and astigmatism. In this work, we assumed corneal astigmatism only and modified the corneal surface property of the Atchison's model. This simple assumption is valid in this case because the prescription lens is only affected by the sum of astigmatism, not the source. The cornea surface profile $C_{rv}$ and $C_{rh}$ are calculated from the cylindrical power, CYL, and AXIS value, and the modified eye model is achieved with SPH value as shown in Table \ref{tab:eyemodel}.

\begin{table}[htb]
    \caption{The modified myopia eye model based on Atchison's model where the $r_x$ and $r_y$ are the radius value of bifocal system in horizontal and vertical respectively, $k_x$ and $k_y$ are the conic constant of bifocal system in horizontal and vertical respectively, $N_d$ is the reflective index of material, and $V_d$ is the Abbe number of material. Note that the radius of cornea surface, $r_x^*$, is calculated by adding the CYL power into another direction as $D_x$ = $D_y$ + CYL, where the $D_y$ = (Nd-1)/$r_y$. The complete equation of $r_x$ is expressed as Eq. (1). }
\resizebox{\textwidth}{!}{
    \centering
    \newcommand{\tabincell}[2]{\begin{tabular}{@{}#1@{}}#2\end{tabular}}
    \begin{tabular}{|c|c|m{4.5cm}<\centering|m{3.5cm}<{\centering}|c|m{3.5cm}<{\centering}|c|}
        \hline
        \textbf{Surface} & \textbf{Type} & \textbf{Radius} & \textbf{Conic} & \textbf{Thickness} & \textbf{Material} & \textbf{Rotation} \\
        \hline
        Cornea & biconic & \multicolumn{1}{|m{4.5cm}|}{$r_x^*$ = (Nd-1)/$D_x$ \par $r_y$ = 7.77+0.022$SR$} & \multicolumn{1}{|m{3.5cm}|}{$k_x$ = -0.15 \par $k_y$ = -0.15} & 0.55 & \multicolumn{1}{|m{3.5cm}|}{Nd = 1.376 \par Vd = 55.468} & 90-AXIS ($^\circ$) \\
        \hline
        Aqueous & standard & 6.40 & -0.275 & 3.05 & \multicolumn{1}{|m{3.5cm}|}{Nd = 1.3337 \par Vd = 50.522} & - \\
        \hline
        Stop & standard & infinite & - & 0.1 & \multicolumn{1}{|m{3.5cm}|}{Nd = 1.337 \par Vd = 50.522} & - \\
        \hline
        Anterior lens & gradient lens & 11.48 & -5.00 & 1.44 & \multicolumn{1}{|m{3.5cm}|}{1.371+0.0652778$Z$ \par -0.0226659$Z^2$ \par -0.0020399($X^2$+$Y^2$)} & - \\
        \hline
        Posterior lens & gradient lens & infinite & - & 2.16 & \multicolumn{1}{|m{3.5cm}|}{1.418-0.0100737$Z^2$ \par -0.0020399($X^2$+$Y^2$)} & - \\
        \hline
        Vitrous & standard & -5.90 & -2.00 & 16.28-0.299$SR$ & \multicolumn{1}{|m{3.5cm}|}{Nd = 1.336 \par Vd = 51.293} & - \\
        \hline
        Retina & biconic & \multicolumn{1}{|m{4.5cm}|}{$r_x$ = -12.91-0.094$SR$ \par $r_y$ = -12.72+0.004$SR$} & \multicolumn{1}{|m{3.5cm}|}{$k_x$ = 0.7+0.026$SR$ \par $k_y$ = 0.225+0.017$SR$} & - & - & - \\
        \hline
        
    \end{tabular}}
    \label{tab:eyemodel}
\end{table}

\begin{equation}
    r_x^* = \frac{r_y \times (Nd-1)}{ (Nd-1) + CYL \times r_y}
\end{equation}

\begin{figure}[hhhh]
\centering
\includegraphics[width=3.5in]{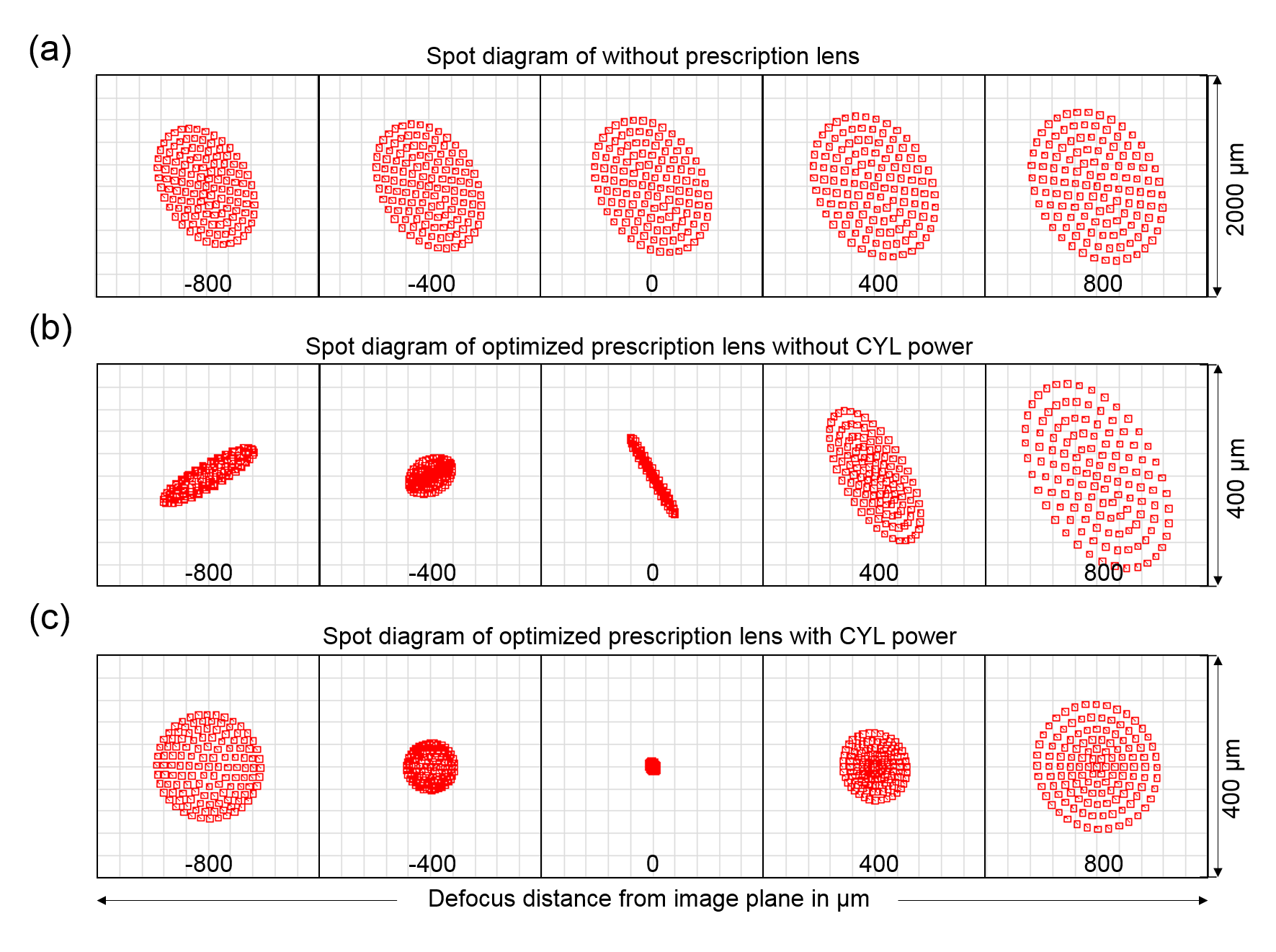}
\caption{\textmd{Analysis of the spot diagram on the retina through focus shifting: (a) the blurred spot on the retina from the infinite object without a prescription lens. (b)The focused spot on the retina only optimized for SPH, and (c) optimized for SPH, CYL, and AXIS. Note that there is serious astigmatism even the spherical optical power is compensated in (b).}}
\label{fig:Prescription lens example}
\end{figure}

Based on this modified myopia eye model, we calculated $S_f$ and $S_r$. $S_f$ is set as a spherical surface of radius $r_f$ while $S_r$ is set as a bifocal surface of radii $r_{ro}$, $r_{re}$ and rotation angle $\theta_r$, to correct the myopia and the astigmatism. All the values were optimized iteratively with the merit function for the range of 12 to 20 mm eye relief, $d_e$, and 26$^\circ$ $\times$ 18$^\circ$ of the field. The Fig. \ref{fig:Prescription lens example} shows the spot diagram change axially around the retina plane of myopic astigmatism eye (SPH: -2, CYL: -2, and AXIS: 30) without and with the prescription lens. Compared to the naked eye focusing at infinite object in Fig. \ref{fig:Prescription lens example}(a), and the myopia-only correction lens in Fig. \ref{fig:Prescription lens example}(b), the designed prescription lens forms a smaller focal point at the retinal plane in Fig. \ref{fig:Prescription lens example}(c).

\subsection{Prescription-embedded AR display design}
\label{sec:AR display design}
Based on the prescription lens design, all other geometric parameters ($R_{l1}$, $R_{l2}$, $R_{cy}$, $a$, $d_p$, $t_{BSL}$, $t_c$, $t_w$, $\theta_d$, $\theta_l$, $\theta_p$, $\theta_c$, $\theta_w$, $\theta_{free}$, and $S_{free}$) were optimized in the second step. Although actual numbers are calculated by Zemax OpticStudio, the geometry of optics, the materials, the constraints should be carefully considered at the design stage for the optimal performance\cite{Extendpoly}. 

\subsubsection{Geometry creation and materials}
There have been several research efforts on free-form image combiner based AR displays \cite{hua20143d,cheng2009design,cheng2011design,wilson2019design} and on general free-form optimization methodology \cite{Freeform_zemax, freeform_nature}. We designed a free-form image combiner for the best imaging quality and the compensation of highly compact off-axis optical design. Figure \ref{fig:schematic diagram of prescription AR}(b) shows the detailed diagram of the AR display path. In the wave-guide, the light rays are reflected at the positive power surface ($S_f$) first, and at the negative power surface ($S_r$) second. So it is reasonable to choose positive power image combiner ($S_{free}$) for the free-form surface for the flatter focal plane, symmetric power distribution, and less lens aberration. The free-form surface can be characterized by an extended polynomial equation including conic aspherical surfaces and extended polynomial terms as follows\cite{Extendpoly}:
\begin{equation}
z=\frac{cr^2}{1+\sqrt{1-(1+k)c^2r^2)}}+\sum_{i}^{N}A_{i}E_{i}(x,y),
\end{equation}
where $c$ is the curvature for the base sphere, $r$ is the normal radius expressed as $r = \sqrt{x^2+y^2}$, $k$ is the conic constant, $N$ is the number of polynomial terms, and $A_{i}$ is the coefficient of the $i^{th}$ extended polynomial terms\cite{Extendpoly}. In our optimization, up to $4^{th}$ polynomials were considered ($N$ = 14).

The bi-convex beam shaping lens increases the system's numerical aperture (NA) for higher resolution and compactness of the shorter optical path. The in-coupling prism guides the light rays into the wave-guide with the TIR condition. The y-axis only cylindrical surface ($R_{cy}$) inside the in-coupling prism compensates some astigmatism and the tilted image plane, which are caused by the off-axis folded path. The tilted angle of the beam shaping lens is identical to the tilted angle of the micro-display for the symmetric magnification ($\theta_d$ = $\theta_l$), but the angles of other components were freely decided by the optimizer to maximize FOV and minimize the aberration. The materials for the beam shaping lens and the upper part of the in-coupling prism ($n_1$, $v_1$), the lower part of that ($n_2$,$v_2$), and the prescription lens ($n_3$,$v_3$), where $n$ and $v$ refer to index of refraction and Abbe number respectively, were carefully chosen to minimize the thicknesses and the chromatic aberration using the different dispersion characteristics. The distances ($a$, $d_p$, $t_{BSL}$, $t_c$, $t_w$) were calculated to some non-negative values based on the following constraints and the priorities.

\subsubsection{Physical Constraints}
The optical configuration for the AR system is limited by giving the constraints in the Merit function. These constraints are determined by the comprehensive consideration of lens implementation, distance from the forehead, total internal reflection condition, and boundary on display panel. In detail, the center thickness of each lens, $t_{BSL}$, $t_c$, $t_w$, and edge thickness must be over 1 $mm$ for manufacturability. The gaps between optical component, $a$ and $d_p$, should be longer than 0.2 $mm$ to avoid overlapping. The sum of thicknesses $a$, $d_p$, $t_{BSL}$, $t_c$, and $t_w$ are limited up to 8.5 $mm$ to minimize the total thickness of AR system so the micro-display can be located in front of the forehead. The light rays used in the optimization were started from the object and limited to the size of the micro display as the final surface.

\subsection{Design trade space}
\label{sec:Design trade space}
\begin{figure}[htb]
\centering
\includegraphics[width=5in]{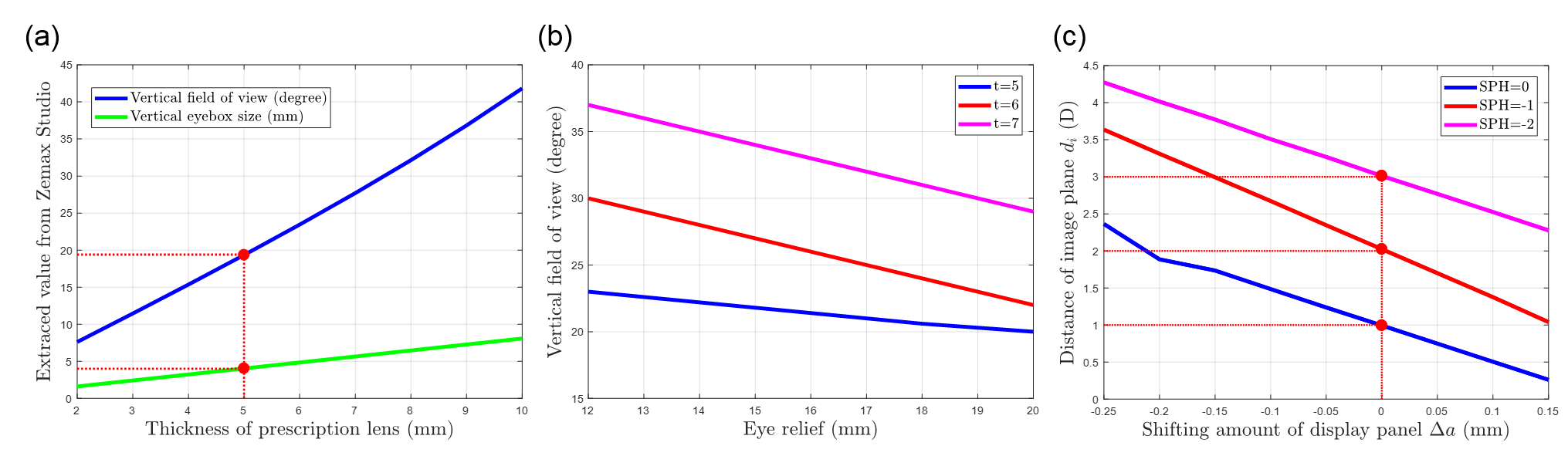}
\caption{\textmd{Design trade-off space for the Prescription AR. The micro display $w_d$$\times$$h_d$ = 10.08 $\times$ 7.56 mm and pixel pitch 6.3 $\mu m$, virtual image plane $d_i = (X+1)D$, and thickness $t_l$ = 5 mm correspond to the ones in the prototype (red circles), where $X$ represents the diopter of myopia. (a) Thickness vs. FOV and eye box. Both FOV and eye box are propotional to the $t_l$. (b) Eye Relief vs. FOV. Smaller eye relief provides larger FOV. (c) Focus cue change. The virtual image plane, $d_i$, can be changed with the axial movement of the micro display, $\Delta a$.}}
\label{fig:Trade_space}
\end{figure}

\subsubsection{FOV vs Thickness and Eye box vs Thickness}
The FOV and the thickness of the prescription lens has a trade-off relation as shown in Fig. \ref{fig:Trade_space}(a).  Especially, a thicker prescription lens allows a larger vertical FOV because of the larger size of the free-form combiner. We chose 5 mm thickness in our design for a comparable FOV with a slim and lightweight form factor. Similarly, the eye box is also decided by the thickness; the thicker prescription lens the larger eye box. The coupling between thickness and display characteristics is reasonable since the height of the micro display $h_d$ is limited by the thickness of the prescription lens. Note that the horizontal FOV and horizontal eye box is determined by the width of the micro display $w_d$ and can be enlarged by choosing long rectangular displays.

\subsubsection{FOV vs Eye relief}
Generally, the more extended eye relief causes the smaller FOV of the near-eye display. Due to the embedded-prescription, the Prescription AR can provide a smaller eye relief (12$\sim$20 mm) compared to the commercially available AR displays, and a comparable FOV with a 0.6-inch display. Since the minimum achievable eye relief might vary per user based on the facial structure, we optimized the optics for 5 different eye relief cases (12, 14, 16, 18 and 20 mm) using the multi-configuration function. The 5-mm thickness prototype can provide 20$\sim$23 degrees vertical FOV based on the user's eye relief.

\subsubsection{Focus cue change}
In addition, Prescription AR provides focal cues. The virtual image plane can be changed by moving the micro display back and forth. The neutral position of the micro display ($a$) was optimized for a 1D focal plane in the vision-corrected scene, which is $(X+1)$D in the real scene for $X$ Diopter myopia (Figure \ref{fig:Trade_space}(c)). By axially shifting the micro display around the original position ($\Delta a$), the virtual image plane can be controlled. Due to the compact optical structure, the overall magnification is large and the depth range from 0D to 2.5D in the vision-corrected scene can be covered with the 0.4 mm axial movement of the micro display. This varifocal characteristic is similarly valid for different myopia levels.

\begin{figure}[hhhh]
\centering
\includegraphics[width=5in]{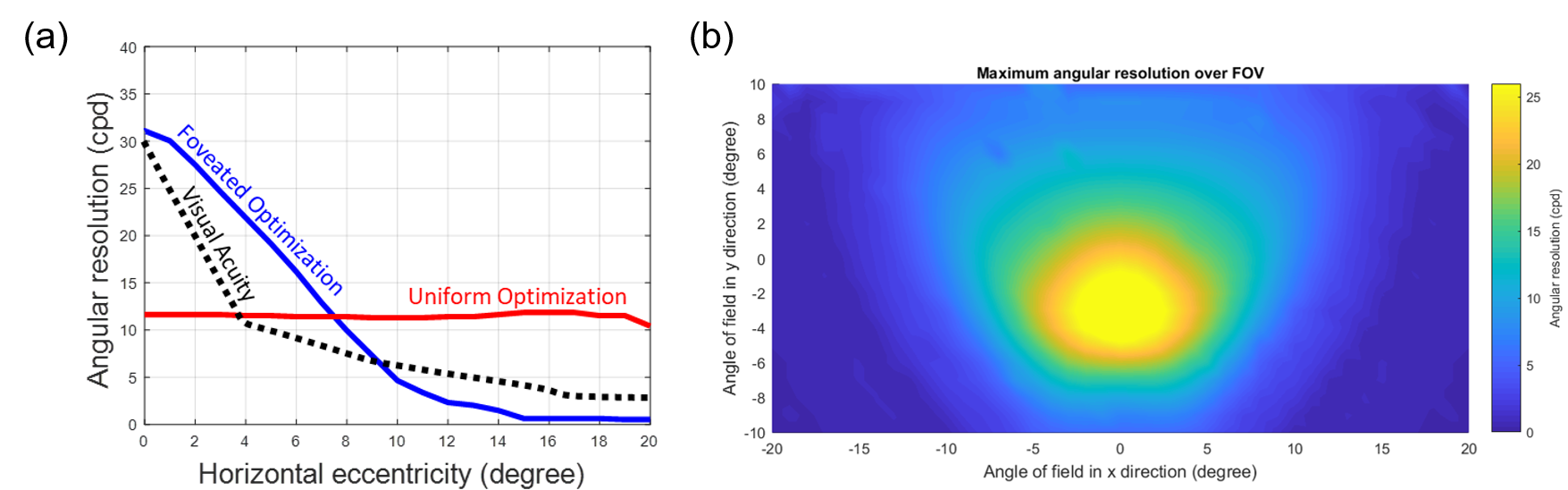}
\caption{\textmd{Foveated optimization of angular resolution in Prescription AR: (a) Human visual acuity over eccentricity (black dotted line), Prescription AR design results (for 1D myopia) with the foveated optimization (blue line), and uniform optimization(red line). Note that each component and the free-form image combiner was optimized for the human visual acuity (fixed foveation) to minimize the wasted information. (b) Angular resolution over FOV in 1D myopia Prescription AR prototype.}}
\label{fig:Angular Resolution SR1}
\end{figure}

\subsubsection{Resolution: foveated optimization}
Because today's micro display pixel numbers are not sufficient to provide 20/20 visual acuity (30 cpd) angular resolution over the full FOV, the foveated display was introduced which distributes the pixels in accordance with the human visual acuity \cite{kim2019foveated, patney2016towards}. Since our display is customized for each user which allows the precise alignment of the eye pupil and the display center, we adopted the foveated optimization. Figure.\ref{fig:Angular Resolution SR1}(a) shows a comparison between a psychophysical visual acuity data and the angular resolution with the uniform and the foveated optimization as a function of horizontal eccentricity \cite{anstis1974chart}. Although uniform optimization can achieve higher resolution in the peripheral region, the user cannot perceive most of the information due to the bandwidth limit. The foveated optimization provides 20/20 visual acuity resolution at the center. Since this system provides fixed foveation, the foveal region and the highest resolution region become misaligned as the eye rotates. Therefore, we sacrificed the resolution of the periphery more and secured higher resolution around the fovea to cover some gaze angle change. The user can perceive 20 cpd resolution image at the 5 degrees gaze angle. The optimized resolution over the FOV is illustrated in Figure \ref{fig:Angular Resolution SR1}(b). The Angular resolution for the vertical eccentricity was optimized at the slightly lower gaze because people tend to look down naturally.

\subsubsection{Presbyopia and hyperopia}
The presbyopia is usually corrected by a bi-focal lens or a progressive lens, which adds an optical power to the lower gaze, for close distance. Since the Prescription AR design only utilizes the upper half of the main lens as a display path, the bi-focal or progressive solution can be directly applied to the current design. For hyperopia, we were not able to find the proper hyperopia eye model. So we directly calculated the prescription lens profiles ($S_f$, $S_r$), and performed the same optimization. The optimized result for 1D hyperopia is shown in Supplementary. The resultant FOV was smaller in 1D hyperopia Prescription AR display because the prescription lens should have a positive optical power and it is difficult to maintain TIR condition. Other than that, the Prescription AR display worked well also in the hyperopia case.

\section{System Implementation}
The Prescription AR can cover most of the myopia, astigmatism, presbyopia, hyperopia, and any combinations of those. We designed Prescription AR for a normal vision case (0D), multiple myopia cases (SPH = -1D, -2D, -3D, -4D, and -5D), a hyperopia case (SPH = 1D) and a myopic astigmatism case (SPH = -2, CYL = -2, AXIS = 30) (see supplementary). Among those designs, we manufactured one case, 1D myopia, for the verification. Table \ref{tab:Table_myopia1D} shows the geometric and optical parameters of Prescription AR for 1D myopia. All the optical components including the mold for the free-form image combiner, the in-coupling prism, and the beam shaping lens were fabricated by ILLUCO. Then the glasses frames were customized for each user. The dynamic and static prototypes were demonstrated. Table \ref{tab:prescriptionAR_1D} shows the detailed specifications for 1D myopia Prescription AR prototype.

\begin{table}[h]
    \centering
    \caption{The 1D myopia prototype of prescription embedded AR display for the experiment}
    \begin{tabular}{|c|c|c|}
        \hline
        \textbf{Items} & \textbf{Units} & \textbf{Values}  \\
        \hline
        Prescription &Diopter & -1  \\ 
        Lens thickness & mm & 5 \\
        Eye relief & mm & 20 \\
        Eye box &$mm^2$ & $6\times4$ \\
        Image plane & m, in the corrected scene & 1 \\
        Field of view & degrees & 20 $\times$ 40 \\
        Resolution & cycles per degree (CPD) & 23 at center \\
        \hline
    \end{tabular}
    \label{tab:Table_myopia1D}
\end{table}

\begin{table}[htb]
\caption{The detail information of -1 SPH prescription AR display design, includes the system parameter and the coefficient of free-form combiner.}
\resizebox{\textwidth}{!}{
    \centering
    \begin{tabular}{|c|c|c|c|c|c|c|c|c|c|c|c|c|c|c|c|c|c|c|}
    \hline
    Prescription & $a$ & $\theta_d$ & $R_{l1}$ & $R_{l2}$ & $n_1$ & $t_{BSL}$ & $\theta_l$ & $d_p$ & $\theta_p$ & $t_c$ & $R_{cy}$ & $n_2$ & $\theta_c$ & $t_w$ & $n_3$ & $\theta_w$ & $\theta_f$ & $d_i$ \\
    \hline
    -1SPH & 0.97 & 66.67 & -37.32 & 13.94 & N-LASF31A & 3.25 & 66.67 & 0.42 & 53.17 & 1.64 & -8.86 & N-BK7 & 61.92 & 2.27 & COP & 62 & 60.92 & 485 \\
    \hline
    Prescription & $c$ & $k$ & $N$ & $r$ & $X^1$ & $Y^1$ & $X^2$ & $X^1Y^1$ & $Y^2$ & $X^3$ & $X^2Y^1$ & $X^1Y^2$ & $Y^3$ & $X^4$ & $X^3Y^1$ & $X^2Y^2$ & $X^1Y^3$ & $Y^4$ \\
    \hline
    -1SPH & -276.28 & 0 & 14 & 27.391 & 0 & 0.798 & 10.65 & 0 & 11.235 & 0 & 0 & 0 & -1.275 & -0.695 & 0 & 0 & 0 & 2.34 \\
    \hline
    \end{tabular}}
    \label{tab:prescriptionAR_1D}
\end{table}

\subsection{Hardware Implementation}
\label{sec:Hardware}
\subsubsection{Prescription-embedded free-form image combiner}
\begin{figure}[hhhh]
    \centering
    \includegraphics[width=5in]{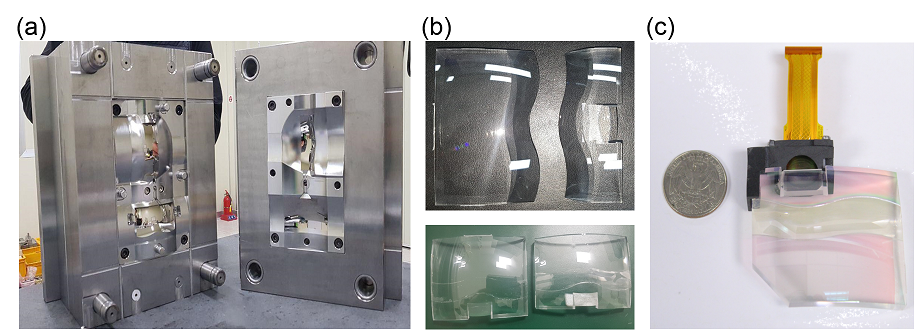}
    \caption{Free-from surface molding process: (a) the mold of free-form surface and waiting for injecting the plastic material, (b) the complete prescription embedded free-form combiner, and (c) the assembled and compact AR display engine.}
    \label{fig:Free_form_mold}
\end{figure}

The fabrication of the customized optical components includes the free-form image combiner are the key for the optical performance. The free-form image combiner was made through the molding processing by ILLUCO. The plastic material, Zeonex cyclo olefin polymer (COP), was used for the prescription lens. It is lightweight and supports the implementation of a free-form surface for molding processing. Figure \ref{fig:Free_form_mold}(a) shows one mold of two lenses since the free-form image combiner separates the prescription lens into upper and bottom lens. The Fig. \ref{fig:Free_form_mold} (b) shows the complete prescription-embedded free-from image combiner after the optics glues are applied on two lenses. The assembled AR display engine is small and lightweight (Fig. \ref{fig:Free_form_mold}(c)).

\subsubsection{Personal Customization}

 \begin{figure}[hhhh]
    \centering
    \includegraphics[width=5in]{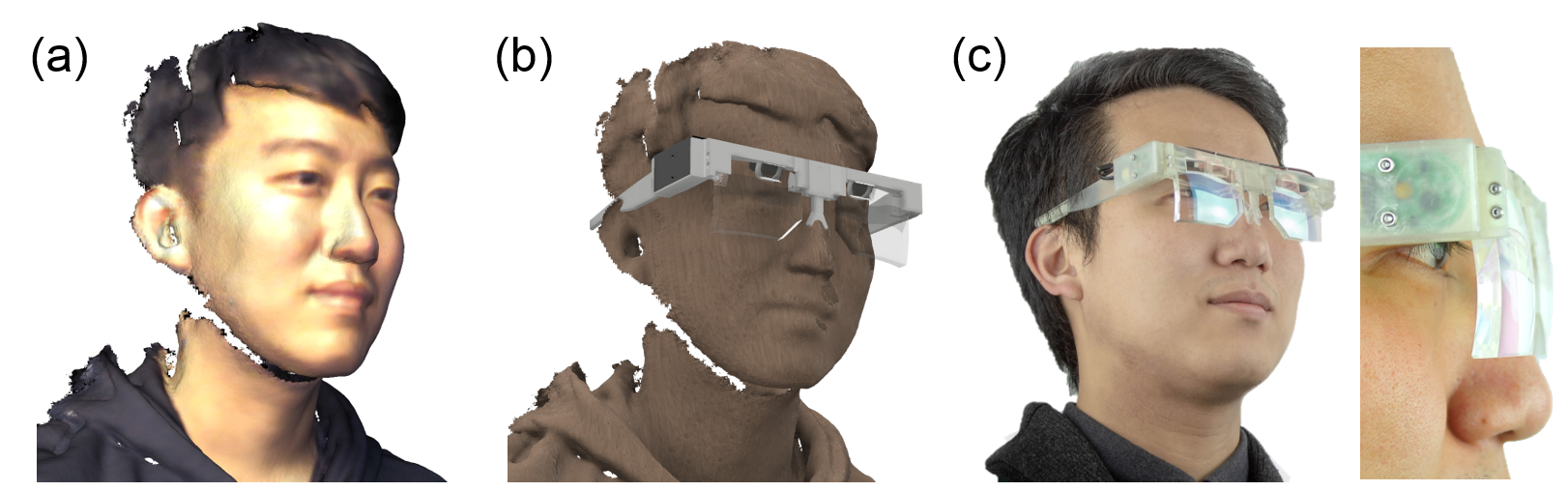}
    \caption{Customized prescription embedded AR display: (a) the scanned 3D model of a user, (b) fitting the glasses frame design with the scanned 3D model, and (c) the results of customization for users.}
    \label{fig:customization}
\end{figure}

 Because every facial structural is unique, the ergonomic frame design is as important as the optics design. Since we optimized the optics for the eye relief of 12 mm to 20 mm, the prototype will work within that range. However, smaller eye relief can provide a larger FOV as Fig. \ref{fig:Trade_space}(b) and more comfortable fit (closer center of mass). Further, the center of the pupil should be aligned with the optical axis for the best foveated experience as Fig. \ref{fig:Angular Resolution SR1}. Therefore, the glasses frame design should consider the user's IPD too.
 
 Figure \ref{fig:customization} shows the customization process of the Prescription AR display. The facial structure was 3D scanned with the Kinect \cite{tong2012scanning,Kinectmicro}, and imported to the 3D rendering software, Fusion 360, where the glasses frame was designed and optimized for each user. The glasses frame designs were parameterized with the input of the IPD and the width of the head, followed by manual fitting for nose part. Figures \ref{fig:customization}(c) show the results of the customization for the different users.

\subsubsection{Prototypes}

\begin{figure}[hhhh]
\centering
\includegraphics[width=5in]{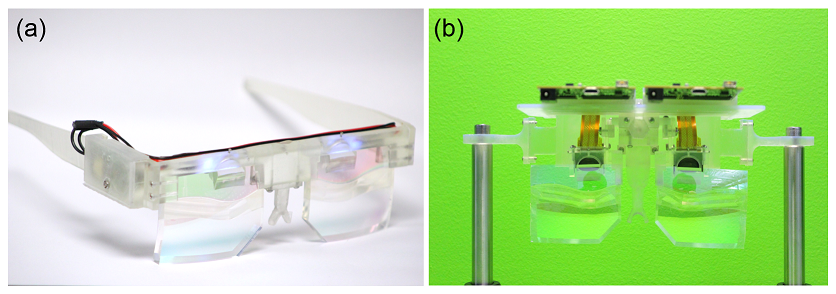}
\caption{\textmd{Implemented prototypes of Prescription AR: (a) static prototype based on LVT films (79 $g$) and (b) dynamic prototype based on micro OLED. 164 $g$ including the driving board.}}
\label{fig:prototype}
\end{figure}

Two Prescription AR prototypes were implemented: a dynamic prototype and a static prototype.

\paragraph{OLED-based dynamic prototype}
Figure \ref{fig:prototype}(a) shows the dynamic prototype. All of the display results were from this setup. Two 10.08$\times$7.56 mm Sony micro OLED (ECX339A) displays were used as binocular micro displays, where each display has 1600$\times$1200 resolution, 6.3 $\mu m$ pixel pitch, and maximum brightness 1000 $cd/m^2$. The free-form optics with the 70\% transparency for 1D myopia were fabricated by Illuco. 3D printed frame housed and aligned all of the optical structures including the main lens attached with an in-coupling prism, beam shaping lens, micro display, and driving board. A 3D printed gear was also applied to change the IPD. The weight of the dynamic prototype including the driving board was 164 $g$.

\paragraph{LVT-based static prototype}
Figure \ref{fig:prototype}(b) shows the static prototype. Two sets of a 10.08$\times$7.56 mm, 3048 pixel per inch light valve technology (LVT) film with an Electro-Luminescent (EL) film backlight were used for the static display. A CR-2032 coin cell powered both EL films. A 3D printed housing aligned all of the optics, statics display modules, and the battery for wearable eyeglasses form factor. The weight of the static prototype was 79$g$.

\subsection{Software Implementation}
\label{sec:Software}
We used a C++ open source innovation engine, called G3D, for rendering \cite{G3D17}. The binocular images with 1200 $\times$ 1600 resolution were rendered in real-time. Over 100 binocular frames were generated every second in a local personal computer with a NVIDIA RTX 2080Ti. The FOV, DOF, IPD were selected carefully in accordance with the actual prototype. Since the resolution was optimized with foveation, the foveated rendering can be considered. We did not apply foveated rendering in this work, but it can be directly applied to the current prototype for reduced computation.
\label{sec:Implementation}

\section{Display Assessment}
\label{sec:Assessment}
\begin{figure}[hhhh]
\centering
\includegraphics[width=3.5in]{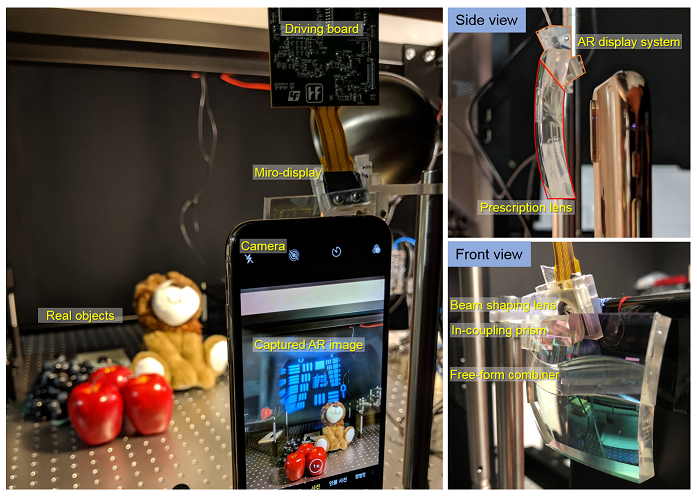}
\caption{\textmd{Photograph of the experimental setup. (left) setup for the experimental results. Note that both the real scene and the virtual image (resolution target) are clearly observed on the phone screen in the normal light condition. (right) close-up photos at the top view and side view.}}
\label{fig:optical bench}
\end{figure}

To properly illustrate the performance of the prototypes of Prescription AR, the experimental results are presented with photographs and videos by an iPhone X camera with the F-number of 2.4 and a 3024 x 4032 resolution. The detailed experimental setup with the camera is shown in Fig. \ref{fig:optical bench}.

\paragraph{Field of view and angular resolution}
\begin{figure}[hhhh]
\centering
\includegraphics[width=5in]{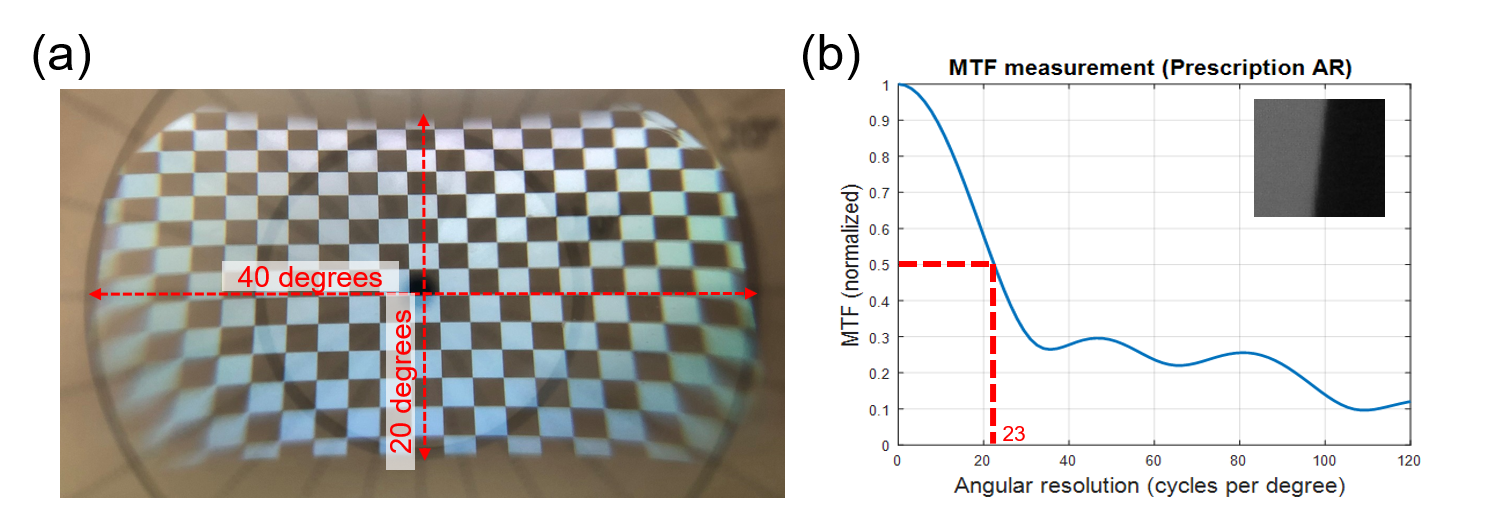}
\caption{\textmd{Experimental results of the FOV and resolution measurement: (a) The AR image covers 20$^\circ$ by 40$^\circ$ of the checking board in vertical and horizontal direction respectively. (b) The analysis of slant-edge measurement at center field. Note that the angular resolution of 23 CPD is realized in our prototype.}}
\label{fig:measurement}
\end{figure}

To measure the FOV, the entire display was illuminated with a white image and was captured with the background FOV panel located at a 15 cm distance as shown in Fig. \ref{fig:measurement}(a). The measured FOV of the prototype was 40$^\circ$ $\times$ 20$^\circ$. The angular resolution is measured from the slant-edge MTF method \cite{burns2000slanted}. Figure \ref{fig:measurement}(b) shows the measured MTF graph and a close-up photo of a slanted edge. The normalized MTF of the Prescription AR was greater than 0.5 at 23 cpd at the center, which was identical to the angular resolution calculated from the pixel pitch (Table \ref{tab:Table_myopia1D}).

\paragraph{Eye-box}
\begin{figure}[hhhh]
\centering
\includegraphics[width=5in]{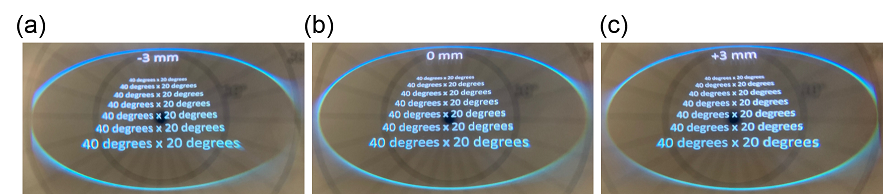}
\caption{\textmd{Eye box measurement results: The $40^\circ \times 20^\circ$ image captured at (a) the left-most, (b) the center, (c) the right-most position. Note that the FOV panel was captured together to show the FOV. The measured eye box was 6$\times$4 $mm^2$.}}
\label{fig:Eye box result}
\end{figure}

Figure \ref{fig:Eye box result} shows the experimental results of the eye box. The full FOV image was displayed and the camera was shifted at the pupil plane to measure the eye box. The prescription AR prototype provided 6 mm of horizontal and 4 mm of vertical eye box while preserving the $40^\circ \times 20^\circ$ FOV.

\paragraph{AR display results with the prescription}
\begin{figure}[hhhh]
\centering
\includegraphics[width=5in]{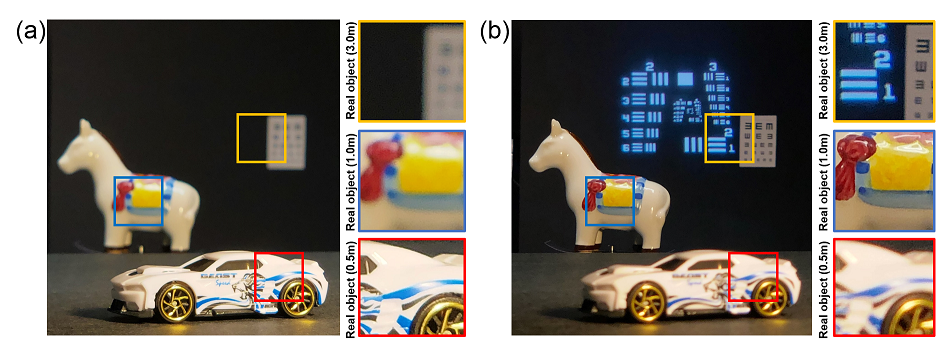}
\caption{\textmd{Experimental results of Prescription AR: (a) Captured image without Prescription AR display. The myopic eye is mimicked as a 1D myopic eye with a fixed focus camera and (b) Captured image with the Prescription AR display. Note that the far object (eye chart) looks sharper with the Prescription AR as well as the virtual image is observed.}}
\label{fig:Prescription result}
\end{figure}

\begin{figure}[hhhh]
\centering
\includegraphics[width=5in]{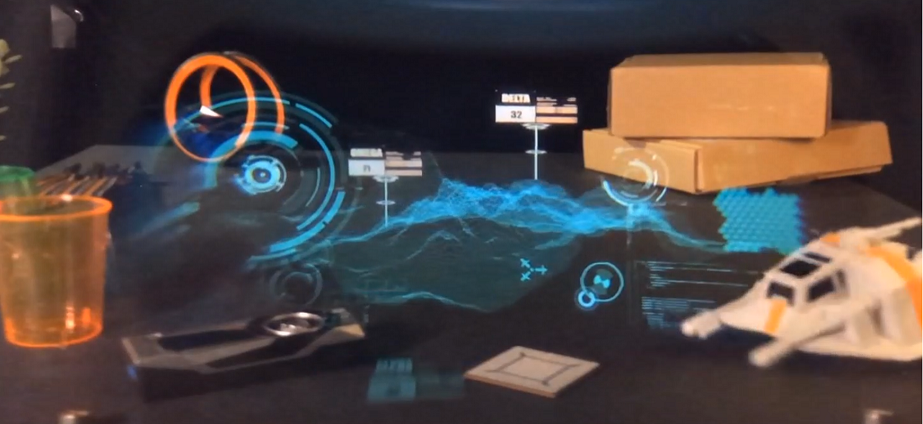}
\caption{\textmd{Experiment video results with the rendered augmented contents generated by rendering software (see Visualization 1)}}
\label{fig:video}
\end{figure}

Figure \ref{fig:Prescription result} shows the captured images without and with the Prescription AR prototype to show the vision-correction results. The real objects were located at different depths: a car, a horse doll, and an eye chart were located at 0.5, 1, and 3 $m$ from the camera respectively in the real world. A fixed focus camera mimicked a 1D myopia eye: the camera was focused at 0.5 $m$ (2D), which is the middle of the 1D myopic eye's DOF. Without the Prescription AR, the camera was focused at the car, was able to resolve some of the details of the horse doll, and was not able to see the details of the eye chart. With the prescription AR, the camera focus was shifted to the horse doll (1D), and was able to achieve good resolution from both of the car and the eye chart. In the real myopic eye where the user can change accommodation, the observer can focus on all three objects with the Prescription AR. The virtual image located at the 1 $m$ distance was also clearly observed in Fig. \ref{fig:Prescription result}(b). Figure \ref{fig:video} shows the video with the rendered augmented contents generated by our rendering software. Note that the virtual image is fixed-focus at the given micro display position.

\paragraph{Focus cue change}
\begin{figure}[h]
\centering
\includegraphics[width=4in]{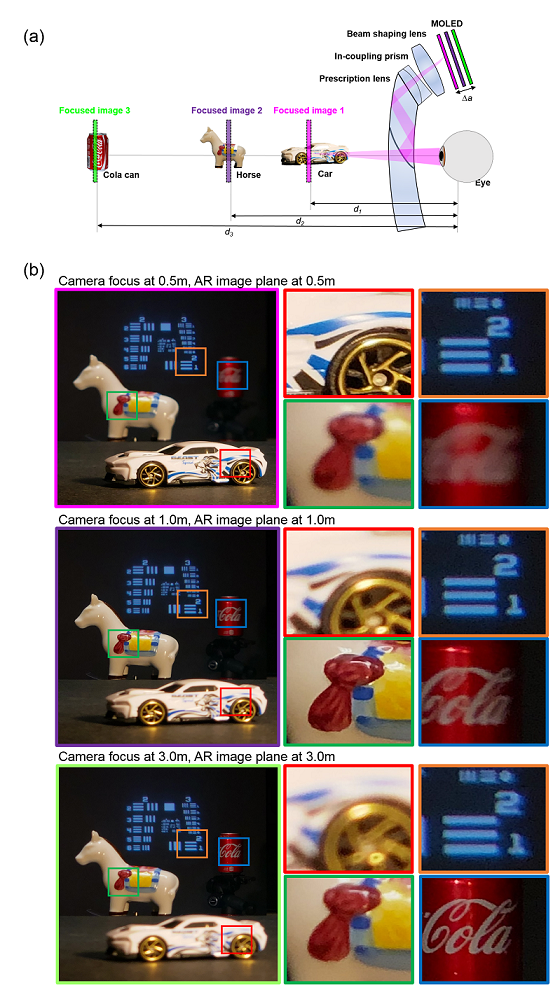}
\caption{\textmd{Focus cue change results: (a)schematic diagram of the experimental setup. Three objects were placed at 0.5 $m$, 1 $m$, and 3 $m$ respectively which related to three virtual images from the certain position of MD with 0.3 $mm$ of shifting, $\Delta a$. (b) the captured image with camera focus at 0.5 $m$ (top), 1 $m$(middle), and 3 $m$(bottom). The clear AR images were perceived at 0.5 $m$ (top), 1 $m$(middle), and 3 $m$(bottom) by corrected position of MD. The prototype can cover from 0.5 $m$ to 3 $m$ by shifting the micro display for 0.4 mm.}}
\label{fig:Focus changing result}
\end{figure}

The virtual image plane is adjustable by moving the micro display back and forth ($\Delta a$). Figure \ref{fig:Focus changing result} shows the experimental results of focus cue change. A car, a horse coll, and a can were located at 0.5 ($d_1$), 1 ($d_2$), and 3 $m$($d_3$) respectively, as shown in Fig. \ref{fig:Focus changing result}(a). The micro display was located accordingly to provide the focus cue. As shown in Fig. \ref{fig:Focus changing result}(b), the in-focus virtual images were observed at 0.5, 1, and 3 $m$ by moving the micro display. The required traveling distance of the micro display was 0.27 $mm$ to cover the depth range from 0.33D to 2D, so the Prescription AR can be free from VAC issue with the submillimeter traveling of the lightweight micro display (< 3$g$).

\paragraph{Brightness and transparency: Privacy and eye-contact interaction}
\begin{figure}[hhhh]
\centering
\includegraphics[width=3in]{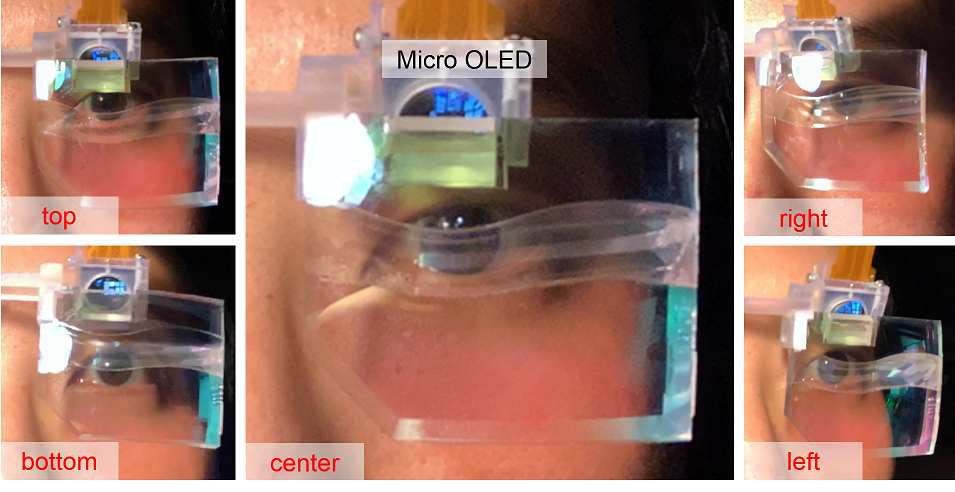}
\caption{\textmd{The captured image of the prototype with the human model. The eye was clearly captured from different at the top, bottom, center, right, and left viewpoint.}}
\label{fig:eyecontact}
\end{figure}

Another essential feature of the AR display is privacy and eye-contact interaction. The privacy is the ability to hide the user's AR contents from outside and the eye-contact interaction is the capability to observe the user's eye from outside while the user is watching AR contents. These issues are closely related to brightness, transparency, and optical configuration. 

The transparency of the Prescription AR prototype was 70\%, and the measured luminance at the pupil plane was 40 $cd/m^2$ while the display brightness was 200 $cd/m^2$. By taking advantages of high light efficiency of the system and the diverging light rays, Prescription AR allows eye-contact interaction while preserving user's privacy. Figure \ref{fig:eyecontact} shows the observed user's face while watching the resolution target contents. The contents were only directly observed from the micro OLED, not through the main lens. Therefore, the user's privacy can be protected by covering the micro OLED region with the non-transparent frames. Furthermore, the user's eye was clearly observed from outside at the different point of views, which enables the eye-contact interaction.

\section{Conclusion}
\label{sec:Conclusion}
 As the internet of things becomes common, the objects around human are getting smarter and interactive. However, unlike other everyday personal items, only the eyeglasses have remained a passive, non-electronic device. Although recent AR display products provide nice augmented experiences with reasonable form factors, they cannot upgrade/substitute eyeglasses without vision-correction. Besides, since people's taste of glasses is very individual, offering a small number pre-designed several options, as true with smartphones, is not a suitable approach for smart glasses business.

In this paper, we showed a fully-customized AR display design approach that considers the user's prescription, IPD, facial structure, and taste of fashion. We established a prescription-embedded AR display optical design method as well as the customization method for individual users. This optical design can cover myopia, hyperopia, astigmatism, and presbyopia, and allows the eye-contact interaction with privacy protection. The glasses and the frames were ergonomically customized using a Kinect-based 3D scanning method. A 169$g$ dynamic prototype showed a 40$^\circ$ $\times$ 20 $^\circ$ virtual image with a 23 cpd resolution at center field and 6 mm $\times$ 4 mm eye box, with the vision-correction and varifocal capability. We believe our work is a start of the paradigm shift in AR display research: from universal design to personalized design. 

\section*{Funding}

\section*{Acknowledgments}
We thank Peter Shirley, Kaan Ak\c{s}it, Zander Majercik, Morgan McGuire, and David Luebke for helpful discussions and advice.

\end{document}